\documentclass[aps,prl,showpacs,superscriptaddress,twocolumn]{revtex4-2}

\usepackage{times,xspace}
\usepackage{graphicx,graphics,color,epsfig}
\usepackage{amsmath}
\usepackage{amssymb}
\usepackage{amsfonts}
\usepackage{amsfonts}
\usepackage{epstopdf}
\usepackage{bm}
\usepackage{hyperref}
\usepackage{color}
\usepackage{soul}
\usepackage{float}
\usepackage[normalem]{ulem}

\def\be{\begin{equation}}
\def\ee{\end{equation}}
\def\ba{\begin{eqnarray}}
\def\ea{\end{eqnarray}}

\newcommand{\figref}[1]{Fig.~\ref{#1}}

\begin{document}

\title{Terahertz harmonic generation across the Mott insulator-metal transition}

\author{G. L. Prajapati}
\email{g.prajapati@hzdr.de}
\affiliation{Institute of Radiation Physics, Helmholtz-Zentrum Dresden-Rossendorf, Bautzner Landstr. 400, 01328 Dresden, Germany}
\author{S. Ray}
\affiliation{Department of Physics, University of Fribourg, 1700 Fribourg, Switzerland}
\author{I. Ilyakov}
\affiliation{Institute of Radiation Physics, Helmholtz-Zentrum Dresden-Rossendorf, Bautzner Landstr. 400, 01328 Dresden, Germany}
\author{A. N. Ponomaryov}
\affiliation{Institute of Radiation Physics, Helmholtz-Zentrum Dresden-Rossendorf, Bautzner Landstr. 400, 01328 Dresden, Germany}
\author{A. Arshad}
\affiliation{Institute of Radiation Physics, Helmholtz-Zentrum Dresden-Rossendorf, Bautzner Landstr. 400, 01328 Dresden, Germany}
\author{T. V. A. G. de Oliveira}
\affiliation{Institute of Radiation Physics, Helmholtz-Zentrum Dresden-Rossendorf, Bautzner Landstr. 400, 01328 Dresden, Germany}
\author{G. Dubey}
\affiliation{Department of Physics, Indian Institute of Science Education and Research Bhopal, Madhya Pradesh 462066, India}
\author{D. S. Rana}
\affiliation{Department of Physics, Indian Institute of Science Education and Research Bhopal, Madhya Pradesh 462066, India}
\author{J.-C. Deinert}
\affiliation{Institute of Radiation Physics, Helmholtz-Zentrum Dresden-Rossendorf, Bautzner Landstr. 400, 01328 Dresden, Germany}
\author{P. Werner}
\affiliation{Department of Physics, University of Fribourg, 1700 Fribourg, Switzerland}
\author{S. Kovalev}
\email{sergey.kovalev@tu-dortmund.de}
\affiliation{Institute of Radiation Physics, Helmholtz-Zentrum Dresden-Rossendorf, Bautzner Landstr. 400, 01328 Dresden, Germany}
\affiliation{Department of Physics, TU Dortmund University, 44227 Dortmund, Germany}

\begin{abstract}
We demonstrate terahertz (THz) harmonic generation across the Mott insulator-metal transition in rare-earth nickelates (RNiO$_3$, R = rare-earth atom). The THz harmonic generation is observed in all the three different phases with distinct behaviors: the intensity of harmonics increases upon cooling in both the low-temperature antiferromagnetic (AFM) insulating and high-temperature paramagnetic (PM) metallic phases, while this trend is reversed in the intermediate PM insulating phase. Using single- and two-band Hubbard models, we find different dominant origins of THz harmonics in different phases: strong spin-charge and orbital-charge couplings in the AFM insulating phase, intraband currents from renormalized quasi-particles with frequency-dependent scattering rate in the PM metallic phase, and the reduction of the charge carrier density due to the opening of the Mott gap in the PM insulating phase. Our study offers strategies for efficient THz harmonic generation from Mott and other strongly correlated systems and insights into the fundamental physics of complex materials. 
\end{abstract}
\vspace{0.5in}

\maketitle
A fundamental manifestation of non-linear light-matter interaction is high harmonic generation (HHG) – the emission of light at integer multiples of the driving frequency. Since the advent of high-power lasers, the number of HHG-based studies on different material systems is continuously growing, starting from atomic and molecular gases to liquids and complex solids \cite{Ferray_1988,Lynga_1996,Luu_2018,Ghimire_2019}. HHG spectra have been extensively used as a probe to investigate the electronic structure of atoms and molecules, the band structure of solids, and inter-atomic bonding \cite{Geneaux_2019,Smirnova_2009,Vampa_2015,Lanin_2017,Luu2_2018}. Further, HHG is the primary method to generate attosecond pulses of frequencies ranging from the extreme ultraviolet to soft x-rays \cite{Krausz_2009,Corkum_2007}. However, these applications have so far been primarily limited to the optical (visible/ultraviolet) regime. Similar activities in the terahertz (THz) regime have started only recently after table-top and electron accelerator-based THz sources with sufficient intensity became available \cite{Kim_2014, Green_2016}. For this reason, the theoretical description of the non-linear optical responses of materials in the THz regime is also not yet fully developed. Nevertheless, within a short span of time, studies based on THz HHG have shown promise in diverse applications, e.g., efficient THz harmonic emission from Dirac and topological materials \cite{Hafez_2018,Kovalev_2020,Cheng_2020,Tielrooij_2022,Kovalev_2021,Ilyakov_Nano,Arshad_2023}, identifying dynamics of Higgs and charge-density wave modes in superconductors \cite{Matsunaga_2014,Chu_2020}, light-wave electronics of the surface states in Dirac materials \cite{Mittendorf_2017}, light field-driven ultrafast spin current generation \cite{Ilyakov_2023}, etc. Here, we report THz HHG study on Mott insulators - an important class of strongly correlated materials that has remained unexplored both theoretically and experimentally. 

Electron-electron correlations play a crucial role in determining the properties of Mott insulators. There are some theoretical and only a few experimental studies investigating the physical processes responsible for optical HHG emission from Mott insulators \cite{Yuta_HHG,Lysne_2020,Murakami_2021,Murakami2_2018,Yuta_CaRuO4, Uchida_Ca2RuO4}. A recent HHG study on the Mott insulator Ca$_2$RuO$_4$ with mid-infrared (MIR) excitation revealed an anomalous temperature-dependence of the HHG intensity: the intensities of higher-order harmonics, whose photon energies are comparable or greater than the Mott gap, increase upon lowering the temperature despite the reduction in charge carrier density \cite{Uchida_Ca2RuO4}. This HHG enhancement was explained on the basis of spin-charge coupling present in the antiferromagnetic (AFM) insulating state \cite{Yuta_CaRuO4}. The same effect was also predicted in the presence of orbital-charge coupling in a paramagnetic (PM) state. These observations raise several interesting questions: (i) can Mott insulators also show HHG upon THz excitation, considering that the THz photon energy (a few meV) is three orders of magnitude smaller than the Mott gap (typically of the order of one eV)? (ii) If yes, what is the mechanism underlying HHG in the THz regime? (iii) Can higher-order THz  harmonics, similar to that observed upon MIR excitation, show anomalous temperature-dependence? Even if they do show anomalous temperature-dependence, we may not observe it experimentally because then we need to detect an order of $100th$ harmonics (to match the energy scale of the Mott gap) and the fact that the intensity of the harmonics usually decreases with increasing its order. (iv) What would be the HHG response in the vicinity of the Mott insulator-metal transition (IMT), i.e., in the correlated metallic state? Above the transition temperature (T$_{\text{IM}}$), HHG in Ca$_2$RuO$_4$ has not been explored. In a similar correlated system, V$_2$O$_3$, HHG was not found in the optical regime above T$_{\text{IM}}$ \cite{Bionta_2021}. Although, THz HHG was recently observed in the quantum critical metal CaRuO$_3$ \cite{Reinhoffer_2024} and in elemental transition metals with nonzero spin-Hall conductivity \cite{Salikhov_arxiv}.

To explore the possibility of THz HHG from Mott insulators and to address the above-mentioned questions, we performed THz third harmonic generation (THG) experiments on rare-earth nickelates (RNiO$_3$, R = rare-earth atom) – a prototype Mott insulator well-known for exhibiting an IMT \cite{Catalano_2018,Middey_2016}. The typical temperature-dependent phase-diagram of nickelates consists of three different regimes: a low temperature antiferromagnetic (AFM) insulating regime, an intermediate temperature paramagnetic (PM) insulating regime and a high temperature PM metallic regime. Not only did the nickelates exhibit a clear THz THG signal in all the three regimes, but also showed distinctly different responses in the different regimes of the phase diagram. Remarkably, the THG also exhibits an anomalous temperature-dependence, i.e, strong THG enhancement upon lowering the temperature in the AFM-insulating regime. Our simulations of the experimental results using single and multi-band Hubbard models reproduce the characteristic features of the THz HHG from Mott insulators. It demonstrates two main points: (i) Thz HHG emission from Mott insulators is indeed possible, and (ii) in contrast to the optical HHG, the enhancement in THz in the AFM insulating phase occurs for all orders of harmonics. This means that, THz HHG does not require photon energies comparable to the width of the Mott gap. Further, the strong sensitivity of the THz HHG spectra on low-energy many-body interactions \cite{Lysne_2020,Yuta_CaRuO4,Murakami_2024} suggests that they can probe fundamental processes in complex systems and provide guidance for future technological applications. 
\begin{figure}
    \centering
    \includegraphics[width=1\linewidth]{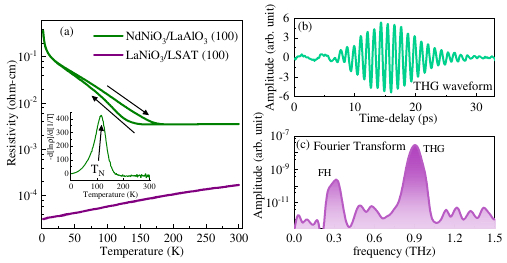}
    \caption{(a) Temperature-dependent resistivity of LaNiO$_3$ and NdNiO$_3$ films. The arrows show the heating and cooling cycles. The inset shows $T_\text{N}$ of the NdNiO$_3$ film determined by the peak in $-\frac{d[\ln{\rho}]}{d[1/T]}$. (b) THG temporal waveform measured after two $0.9$ THz bandpass filters for the LaNiO$_3$ film at $T=22$~K. Its Fourier transform is shown in (c). The peaks at $0.3$ THz and $0.9$ THz represent the driving pulse and THG, respectively.}
    \label{fig_1}
\end{figure}

Two high quality films of LaNiO$_3$ and NdNiO$_3$ were fabricated on LaAlO$_3$ ($100$) and LSAT ($100$) single crystal substrates, respectively (for detailed thin film growth and structural characterization, see Supplementary Material (SM) \cite{sup_mat}). LaNiO$_3$/LAO ($100$) remains a PM metal down to low temperature, while NdNiO$_3$/LSAT ($100$) exhibits an AFM to PM transition at $\sim 106$ K and an IMT at $\sim 150$ K, as inferred from their temperature-dependent resistivity data shown in Fig.~\ref{fig_1}(a) (also see section S3 \cite{sup_mat}). The THz THG measurements on these films were performed using intense narrowband multicycle THz pulses ($E_{\text{THz}} \sim 100$ kV/cm) with central frequency $\omega = 0.3$ THz. The transmitted THz radiation from the films was filtered using two bandpass filters with central frequency $3\omega = 0.9$ THz which suppress significantly the residual radiation at the fundamental frequency. The THG signal was then detected in the time-domain using a 2~mm thick ZnTe crystal (for further details on the experimental setup, see section S4 \cite{sup_mat}). An exemplary THG waveform emitted from the LaNiO$_3$ film is shown in Fig.~\ref{fig_1}(b) and its Fourier transform is plotted in Fig.~\ref{fig_1}(c). The observation of a dominant signal at $0.9$~THz confirms the THG emission from the LaNiO$_3$ film.
\begin{figure}
    \centering
    \includegraphics[width=1\linewidth]{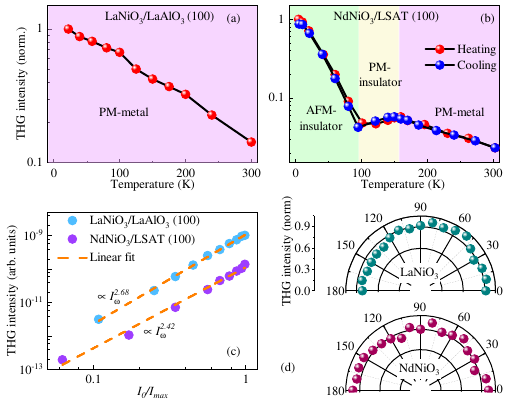}
    \caption{Temperature-dependent THG intensity for (a) LaNiO$_3$ and (b) NdNiO$_3$ films. Different colored regions in (b) represent the temperature-dependent different electronic states of the NdNiO$_3$ film. (c) THG intensity as a function of fundamental THz intensity for LaNiO$_3$ ($T=22$~K) and NdNiO$_3$ ($T=10$~K) films. Dotted lines are linear fits to the experimental data. (d) Polar plots of THG from LaNiO$_3$ and NdNiO$_3$ films at $10$ K as a function of the driving THz pulse polarization angle.}
    \label{fig_2}
\end{figure}

Fig.~\ref{fig_2}(a) shows the temperature-dependent THG intensity ($I_{3\omega}$) of the PM metallic LaNiO$_3$ film. The THG intensity increases monotonously upon lowering temperature up to the lowest measured temperature $T \sim 20$ K. The same holds true for the PM metallic state of the NdNiO$_3$ film (see Fig.~\ref{fig_2}(b)). However, below $T_{\text{IM}}$, the THG intensity starts decreasing and this trend continues down to the magnetic transition temperature ($T_{\text{N}}$). Remarkably, below $T_{\text{N}}$, the THG intensity starts rising sharply and monotonously down to the lowest temperature, showing an increase in THG intensity by more than one order of magnitude at $\sim 5$ K compared to that at $T_{\text{N}}$. Fig.~\ref{fig_2}(c) shows the THG intensity ($I_{3\omega}$) as a function of fundamental THz intensity ($I_{\omega}$) for both films. Ideally, one expects a cubic relation $I_{3\omega} \propto {I_{\omega}}^3$; however, for both films, the exponent is significantly smaller than $3$. This indicates that the THG emission from the nickelate films originates from light-matter interaction in the non-perturbative regime \cite{Cheng_2020}.
Further, one may expect a crystal orientation-dependent THG response due to strain-induced modifications in the magnitude of O-$2p$ and Ni-$3d$ orbital polarizations in different directions \cite{Peng_2016}. To determine the directional dependence of the THG, THz polarization-dependent THG measurements were performed at $T=10$ K for both films. As shown in Fig.~\ref{fig_2}(d), the THG intensity in both cases is nearly independent of the THz polarization orientation. This indicates that the sum of the contributions to the THG signal is isotropic. This is also supported by the fact that the films are isotropically strained due to the cubic structure of the underlying substrates. Consequently, the magnitude of the orbital polarization should be identical along the two orthogonal in-plane directions. Hence, as expected, an isotropic response of the THG signal to the polarization of the THz field is observed.

The HHG originating from light-matter interaction in strongly correlated systems can be simulated with the nonequilibrium DMFT formalism \cite{Markus_1,Markus_2,Yuta_HHG, Yuta_CaRuO4}. HHG in single-band Mott insulators originates both from the intra-Hubbard-band dynamics of doublons (doubly occupied states) and holons (empty states) and from their recombination \cite{Yuta_CaRuO4}. A similar situation can be expected in two-orbital systems in the paramagnetic (PM) state. Here, the hopping between singlons (singly occupied states) and triplons (triply occupied states) however leads to orbital-charge coupling due to the orbital moments of the triplons and singlons \cite{Yuta_CaRuO4}.
The situation in the AFM phase of the nickelates is more complex, since spin-charge and orbital-charge couplings are simultaneously active and their interplay  needs to be considered in the modelling. It is also necessary to choose a sufficiently low excitation frequency in order to interpret the THz HHG experiments, where the photon energy is three-orders of magnitude smaller than the Mott gap. This is numerically challenging, because it requires simulations up to long times. 

\begin{figure*}[t]
\includegraphics[width=0.95\textwidth]{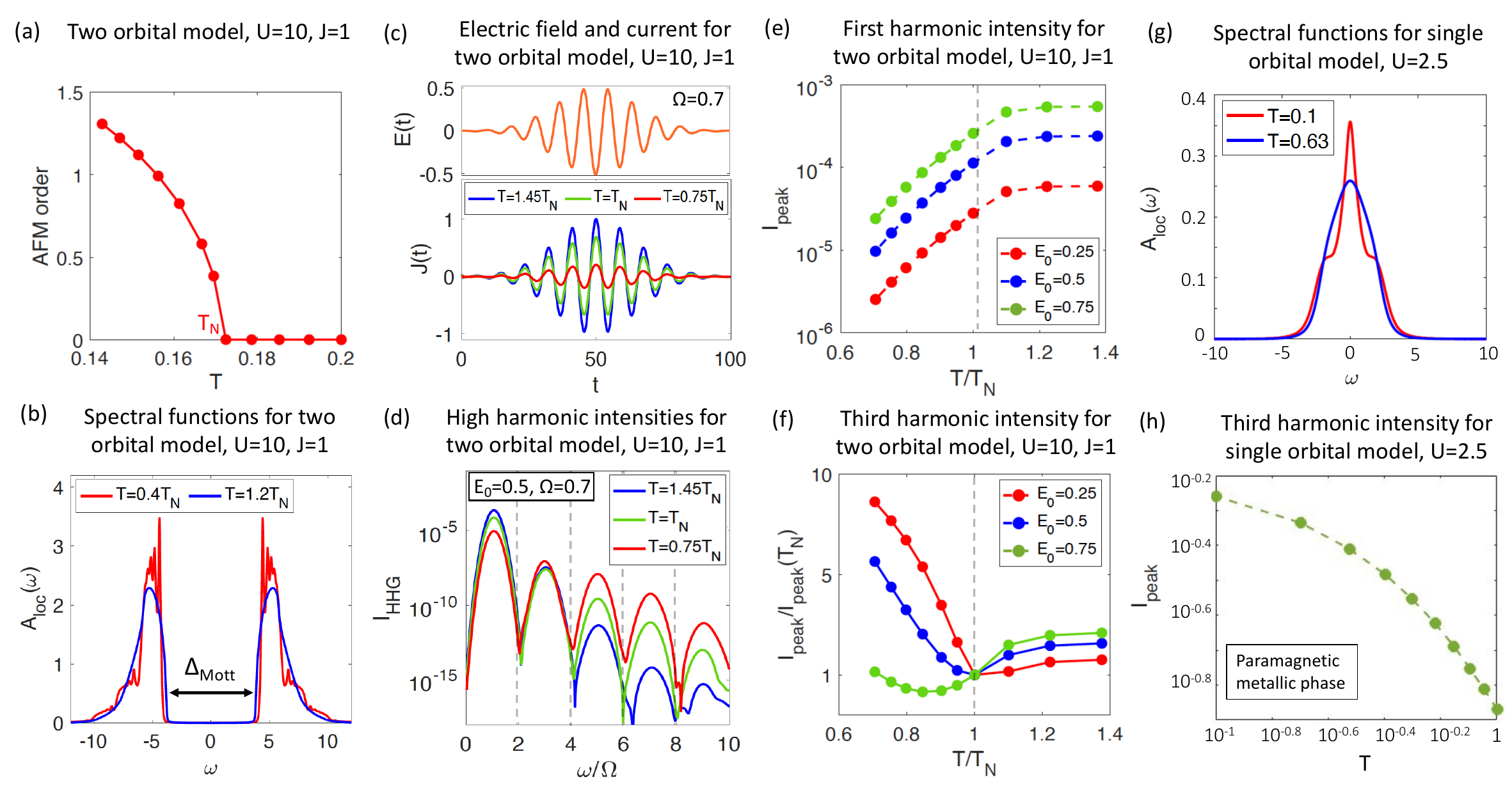}
\caption{ DMFT results for the two-orbital repulsive Hubbard model with $U=10, J=1$. (a) AFM order as a function of temperature with the N\'eel temperature $T_{\text{N}}$ indicated. (b) Single-particle local spectral function $A_{\text{loc}}(\omega)$ below and above $T_{\text{N}}$. (c) Electric field pulse $E(t)$ with $\Omega=0.7$ (upper panel) and the resultant current $\mathcal{J}(t)$ (not to scale) for different temperatures (lower panel). (d) High harmonic intensities for $E_0=0.5$ and $\Omega=0.7$ for different temperatures (obtained from the Fourier transform of the $\mathcal{J}(t)$ shown in (c)). (e) Peak values $I_{\text{peak}}$ of the first harmonic intensity and (f) $I_{\text{peak}}$ of the third harmonic intensity as a function of temperature for different electric field strengths $E_0$. (g) Local spectral function $A_{\text{loc}}(\omega)$ of the single-orbital repulsive Hubbard model with $U=2.5$ for different temperatures. (h) Peak values $I_{\text{peak}}$ of the third harmonic intensity as a function of temperature for the single-orbital Hubbard model with $U=2.5$.}
\label{fig_3}
\end{figure*}

To account for the relevant factors, we use the two-orbital Hubbard model with Hamiltonian
\begin{align}
&H = -t_\text{hop}\sum_{\left<ij\right>,\sigma}\sum_{\alpha=1,2} c^{\dagger}_{i,\alpha\sigma}c_{j,\alpha\sigma} \nonumber + U\sum_{i}\sum_{\alpha=1,2}n_{i,\alpha\uparrow}n_{i,\alpha\downarrow} \\
&\hspace{0mm} - \mu \sum_{i}\sum_{\alpha=1,2}\left(n_{i,\alpha\uparrow} + n_{i,\alpha\downarrow} \right) \nonumber + (U-2J)\sum_{i,\sigma}n_{i,1\sigma}n_{i,2\Bar{\sigma}} \\
&\hspace{0mm} + (U-3J)\sum_{i,\sigma}n_{i,1\sigma}n_{i,2\sigma} ,
\label{eq_1}
\end{align}
where, $\alpha$ and $\sigma$ denote the orbital and spin indices, respectively, $t_\text{hop}$ is the nearest-neighbor hopping amplitude between sites $i$ and $j$, $U$ is the intra-orbital Hubbard repulsion, $J$ the Hund coupling, and $\mu$ the chemical potential. We use a code based on the NESSi library \cite{NESSi} and a non-crossing approximation (NCA) impurity solver \cite{Keiter_1971,Eckstein_2010} to analyze the dynamics of large-gap Mott insulators driven by an electric field pulse $E(t) = E_0f_g(t)\sin(\Omega t)$, where $\Omega$ is the pulse frequency and $f_g$ is a Gaussian envelope function (\figref{fig_3}(c)). In such a driven system, charge carriers are created with a density controlled by the electric field strength $E_0$, and their kinematics are coupled to the correlated background of spin and orbital moments. In a single-orbital system, strong spin-charge coupling of the charge excitations (doublon-holon pairs) with the AFM spin background gives rise to an increase in HHG below the Neel temperature (T$_{\text{N}}$) \cite{Yuta_CaRuO4}. With decreasing temperature, the coherence between the doublons and holons increases because of suppressed thermal fluctuations in the spin-background. This reduces the phase cancellations in the emitted light, enhances the high harmonic intensities, and leads to a more pronounced HHG plateau \cite{Yuta_HHG, Yuta_CaRuO4, Yuta_1Dmott}.

In a two-orbital system, although the basic concept of coupling of the charge excitations to a correlated background remains the same, there are some differences because the background now features both spin and orbital moments. For Mott insulators with positive Hund coupling ($J>0$) the anomalous temperature dependence of the HHG peaks is observed for the fifth and higher harmonics even above the Neel temperature, because the hopping of singlons (singly occupied sites) and triplons (triply occupied sites) in a background of high-spin doublons can leave behind low-spin configurations. If we decrease the temperature further below the N\'eel temperature ($T_\text{N}$), spin-charge coupling with excitation energies of order $J_{\text{ex}} \ll J$ is activated, and we observe an increase in high harmonic intensities with decreasing temperature for the third and higher harmonics. In ~\figref{fig_3}(a) we show the AFM order as a function of temperature, while ~\figref{fig_3}(b) shows the equilibrium single-particle spectral function at different temperatures for the two-orbital Hubbard model with $U=10, J=1$, which are approximately the realistic parameters for nickelates \cite{Ni_Hariki}. Below $T_\text{N}$, the Mott gap increases with decreasing temperature and the Hubbard bands split into sharp subbands due to the formation of spin-polarons. We measure the high harmonic intensity I$_{\text{HHG}}(\omega)$ from the Fourier transform of the current $\mathcal{J}(t)$ as $I_{\text{HHG}}(\omega)=|\omega \mathcal{J}(\omega)|^{2}$. $\mathcal{J}(t)$ is shown for a few temperatures in ~\figref{fig_3}(c), while ~\figref{fig_3}(d) plots $I_{\text{HHG}}(\omega)$ for the two-orbital Hubbard model with $U=10, J=1$ and driving frequency $\Omega=0.7$ for different temperatures. Due to inversion symmetry, we get peaks at odd multiples of the driving frequency $\Omega$. The increase in the intensity of the higher harmonics with decreasing temperature (both above and below $T_{\text{N}}$) is clearly seen.

A different behavior is found for the first and third harmonics. In ~\figref{fig_3}(e) and (f), we plot the peak values of the first and third harmonic intensities, respectively as a function of temperature and at different fixed electric field strengths. We start with $E_0=0.25$ (red curve) and a temperature above $T_\text{N}$ in the PM insulating region. As we gradually decrease the temperature, we observe a decrease in the first and third harmonic intensities. This is due to the increase of the Mott gap with decreasing temperature and the suppression of the density of charge carriers. Once we are below $T_\text{N}$, we observe an increase in the THG intensity with decreasing temperature. In this region, although the number of the induced charge carriers still decreases with decreasing temperature (as reflected in the decreasing first harmonic intensity), there is a strong spin-charge coupling of the charge carriers with the ordered background. The formation of spin-polarons is evident from the appearance of sharp peaks in the single-particle spectral functions at equilibrium below $T_\text{N}$ (see \figref{fig_3}(b)). Next, we increase the electric field strength to $E_0=0.5$ and observe an increase in the overall first harmonic intensity due to the increased density of charge carriers. At the same time, the rate of increase of the THG intensity decreases. If we increase the electric field strength further to $E_0=0.75$, the THG intensity decreases even below the N\'eel temperature but starts increasing below about $0.8 T_{\text{N}}$ (\figref{fig_3}(f)). A higher electric field leads to a stronger disordering of the spin background and we need stronger correlations in the initial state to observe the increase in THG intensity with decreasing temperature, which is achieved at $T \ll T_{\text{N}}$. In this context, it is interesting to note that the experiments in Ref. \cite{Uchida_Ca2RuO4}, which do not observe any THG increase, used a field strength which is more than two-orders of magnitude higher than that in the present study.

In the high-temperature PM metallic phase, the physics can be approximately described by a single-orbital model. To analyze the THG in the metallic region, we use the single-orbital Hubbard model with $U=2.5$ and a one-crossing approximation (OCA) impurity solver, which is computationally more costly than NCA, but provides a better description of metallic states. The local spectral function of this model is shown in ~\figref{fig_3}(g). At high temperatures, the metallic spectrum is relatively broad, but at lower temperatures, a sharp low-energy peak emerges, indicative of coherent quasi-particle excitations. Consistent with this, we observe that the THG intensity increases with decreasing temperature in the metallic phase, as seen in ~\figref{fig_3}(h). The HHG emission in the metallic phase can be attributed to the intraband dynamics of the quasiparticles, which for large field amplitude results in a non-linear response \cite{Reinhoffer_2024}. Due to strong electronic correlations, nickelates often exhibit metallic states with both incoherent and coherent features, indicative of an energy- (or frequency-) dependent scattering rate 
\cite{Phanindra_2018,Prajapati_2021}. Below $T_{\text{IM}}$, the Mott gap opens and increases with decreasing temperature, leading to a suppression of the induced charge carrier density. At sufficiently high temperatures, this effect dominates over the orbital-charge coupling effect on HHG, leading to a decrease in HHG intensity with decreasing temperature in the PM-insulating regime.

In summary, we experimentally demonstrated THz HHG from a prototypical Mott insulator and successfully modeled the observed temperature-dependence of the HHG on the fundamental quasiparticle interactions. The THz THG is visible in all the three different regimes of the nickelate phase diagram along with a remarkably sharp enhancement in the AFM insulating regime. This enhancement occurs due to the combined effect of spin-charge and orbital-charge couplings.   
In the PM correlated metallic state, the HHG mechanism is dominated by the responses of heavy quasiparticles near the Fermi level and their frequency-dependent scattering rate. In the intermediate PM insulating state, the characteristic temperature-dependence of HHG emission is governed by opening of the Mott gap. All these factors responsible for HHG emission in nickelates, can be found in other strongly correlated systems, particularly, in 3d to 5d transition metal oxides \cite{Kumar_2020}. In 4d and 5d transition metal oxides, spin-orbit coupling is also expected to play a relevant role in THz HHG emission \cite{Markus_1}. Our study demonstrates the general feasibility of THz HHG from these and similar systems. As the THz HHG response is highly sensitive to microscopic low-energy degrees of freedom and the corresponding quasiparticle interactions, this method can provide relevant insights into the physics of complex systems. Further, new strategies for improving the efficiency of HHG from correlated systems may be devised, extending the range of efficient HHG materials beyond Dirac and topological materials. Finally, our study demonstrates that recently developed theoretical schemes such as nonequilibrium DMFT can be extended down to the THz range, accurately modeling nonlinear light-matter interactions. Thereby, these models may help to drive the
progress of THz HHG-based science and technology.

Parts of this research were carried out at ELBE at the Helmholtz-Zentrum Dresden-Rossendorf e.V., a member of the Helmholtz Association. S. R. and P. W. acknowledge support from SNSF Grant No. 200021- 196966. J.-C. D. and G. L. P. acknowledge support from BMBF Verbundprojekt 05K2022 - Tera-EXPOSE.


\begin{thebibliography}{99}

\bibitem{Ferray_1988} M. Ferray, A. L’Huillier, X. F. Li, L. A. Lompre, G. Mainfray, and C. Manus, J. Phys. B {\bf 21}, L31 (1988).

\bibitem{Lynga_1996} C. Lyngå, A. L'Huillier, and C-G Wahlström, J. Phys. B: At. Mol. Opt. Phys. {\bf 29}, 3293 (1996).

\bibitem{Luu_2018} T. T. Luu, Z. Yin, A. Jain, T. Gaumnitz, Y. Pertot, J. Ma, and H. J. Wörner, Nat Commun {\bf 9}, 3723 (2018).

\bibitem{Ghimire_2019} S. Ghimire, and D. A. Reis, Nat. Phys. {\bf 15}, 10–16 (2019).

\bibitem{Geneaux_2019}	R. Geneaux, H. J. Marroux, A. Guggenmos, D. M. Neumark, and S. R. Leone, Philos. Trans. R. Soc. A {\bf 377}, 20170463 (2019).

\bibitem{Smirnova_2009} O. Smirnova, Y. Mairesse, S. Patchkovskii, N. Dudovich, D. Villeneuve, P. Corkum, and M. Y. Ivanov, Nature (London) {\bf 460}, 972 (2009).

\bibitem{Vampa_2015} G. Vampa, T. J. Hammond, N. Thiré, B. E. Schmidt, F. Légaré, C. R. McDonald, T. Brabec, D. D. Klug, and P. B. Corkum, Phys. Rev. Lett. {\bf 115}, 193603 (2015).

\bibitem{Lanin_2017} A. A. Lanin, E. A. Stepanov, A. B. Fedotov, and A. M. Zheltikov, Optica {\bf 4}, 516 (2017).

\bibitem{Luu2_2018} T. T. Luu, and H. J. Wörner, Nat. Commun. {\bf 9}, 916 (2018).

\bibitem{Krausz_2009} F. Krausz, M. Ivanov, Rev. Mod. Phys. {\bf 81}, 163-234 (2009).

\bibitem{Corkum_2007} P. B. Corkum, and F. Krausz, Nat. Phys. {\bf 3}, 381-387 (2007).

\bibitem{Kim_2014} K. Y. Kim, and Y. S. You, Nat. Photonics, {\bf 8}, 92-94 (2014).

\bibitem{Green_2016} B. Green et al., Sci. Rep. {\bf 6}, 22256 (2016).

\bibitem{Hafez_2018} H. A. Hafez et al., Nature {\bf 561}, 507-511 (2018).

\bibitem{Kovalev_2020} S. Kovalev et al., Nat. Commun. {\bf 11}, 2451 (2020).

\bibitem{Cheng_2020} B. Cheng, N. Kanda, T. N. Ikeda, T. Matsuda, P. Xia, T. Schumann, S. Stemmer, J. Itatani , N. P. Armitage, and R. Matsunaga, Phys. Rev. Lett. {\bf 124}, 117402 (2020).

\bibitem{Tielrooij_2022} K.-J. Tielrooij et al., Light Sci. Appl. {\bf 11}, 315 (2022).

\bibitem{Kovalev_2021} S. Kovalev et al., npj Quantum Mater. {\bf 6}, 84 (2021).

\bibitem{Ilyakov_Nano} I. Ilyakov et al., Nano Lett. {\bf 23}, 3872–3878 (2023). 

\bibitem{Arshad_2023} A. Arshad et al., Adv. Photonics Res. {\bf 4}, 2300088 (2023). 

\bibitem{Matsunaga_2014} R. Matsunaga, N. Tsuji, H. Fujita, A. Sugioka, K. Makise, Y. Uzawa, H. Terai, Z. Wang, H. Aoki, R. Shimano, Science {\bf 345}, 1145-1149 (2014).

\bibitem{Chu_2020} H. Chu et al., Nat. Commun. {\bf 11}, 1793 (2020).

\bibitem{Mittendorf_2017} M. Mittendorﬀ, S. Li, and T. E. Murphy, ACS Photonics {\bf 4}, 316-321 (2017).

\bibitem{Ilyakov_2023} I. Ilyakov, A. Brataas, T. V. A. G. de Oliveira, A. Ponomaryov, J.-C. Deinert, O. Hellwig, J. Faßbender, J. Lindner, R. Salikhov, and S. Kovalev, Nat. Commun. {\bf 14}, 7010 (2023). 

\bibitem{Yuta_HHG} Y. Murakami, M. Eckstein, and P. Werner, Phys. Rev. Lett. {\bf 121}, 057405 (2018).

\bibitem{Lysne_2020} M. Lysne, Y. Murakami, and P. Werner, Phys. Rev. B {\bf 101}, 195139 (2020).

\bibitem{Murakami_2021} Y. Murakami, S. Takayoshi, A. Koga, and P. Werner, {\bf 103}, 035110 (2021).

\bibitem{Murakami2_2018} Y. Murakami, and P. Werner, Phys. Rev. B {\bf 98}, 075102 (2018).

\bibitem{Yuta_CaRuO4} Y. Murakami, K. Uchida, A. Koga, K. Tanaka, and P. Werner, Phys. Rev. Lett. {\bf 129}, 157401 (2022).

\bibitem{Uchida_Ca2RuO4} K. Uchida, G. Mattoni, S. Yonezawa, F. Nakamura, Y. Maeno, and K. Tanaka, Phys. Rev. Lett. {\bf 128}, 127401 (2022).

\bibitem{Bionta_2021} M. R. Bionta et al., Phys. Rev. Res. {\bf 3}, 023250 (2021).

\bibitem{Reinhoffer_2024} C. Reinhoffer, S. Esser, S. Esser, E. A. Mashkovich, S. Germanskiy, P. Gegenwart, F. Anders, P. H. M. van Loosdrecht, and Z. Wang, Phys. Rev. Lett. {\bf 132}, 196501 (2024).

\bibitem{Salikhov_arxiv} R. Salikhov et al., arXiv:2311.13272.

\bibitem{Catalano_2018} S. Catalano, M. Gibert, J. Fowlie, J. Íñiguez, J.-M. Triscone, and J. Kreisel, Rep. Prog. Phys. {\bf 81}, 046501(2018).

\bibitem{Middey_2016} S. Middey, J. Chakhalian, P. Mahadevan, J. W. Freeland, A. J. Millis, and D. D. Sarma, Annu. Rev. Mater. Res. {\bf 46}, 305-334 (2016).

\bibitem{Murakami_2024} Y. Murakami, T. Hansen, S. Takayoshi, L. B. Madsen, and P. Werner, arXiv:2407.07752 (2024).

\bibitem{sup_mat} Supplemental Material

\bibitem{Peng_2016} J. J. Peng, C. Song, M. Wang, F. Li, B. Cui, G. Y. Wang, P. Yu, and F. Pan, Phys. Rev. B {\bf 93}, 235102 (2016).

\bibitem{Markus_1} M. Lysne, Y. Murakami, and P. Werner, Phs. Rev. B {\bf 101}, 195139 (2020).

\bibitem{Markus_2} M. Lysne, M. Sch\"uler Y. Murakami, and P. Werner, Phs. Rev. B {\bf 102}, 081121(R) (2020).

\bibitem{NESSi} M. Sch\"uler, D. Golelez, Y. Murakami, N. Bittner, A. Herrmann, H. U. R. Strand, P. Werner, and M. Eckstein, Comput. Phys. Commun. {\bf 257}, 107484 (2020).

\bibitem{Keiter_1971} H. Keiter and J. C. Kimball, Int. J. Magn. {\bf 1}, 233 (1971).

\bibitem{Eckstein_2010} M. Eckstein and P. Werner, Phys. Rev. B {\bf 82}, 115115 (2010).

\bibitem{Yuta_1Dmott} Y. Murakami, S. Takayoshi, A. Koga, and P. Werner, Phys. Rev. B {\bf 103}, 035110 (2021). 

\bibitem{Ni_Hariki} A. Hariki, M. Winder, T. Uozumi, and J. Kuneš, Phys. Rev. B {\bf 101}, 115130 (2020).

\bibitem{Phanindra_2018} V. E. Phanindra, P. Agarwal, and D. S. Rana Phys. Rev. Mater. {\bf 2}, 015001 (2018).

\bibitem{Prajapati_2021} G. L. Prajapati, S. Das and D. S. Rana, J. Phys. Condens. Matter {\bf 33}, 415401 (2021).

\bibitem{Kumar_2020} K. S. Kumar, G. L. Prajapati, R. Dagar, M. Vagadia, D. S. Rana, M. Tonouchi, Adv. Optical Mater {\bf 8}, 1900958 (2020).

\end{thebibliography}
\end{document}